\documentclass[aps,prl,twocolumn,superscriptaddress,groupedaddress]{revtex4}  
\usepackage{graphicx}  
\usepackage{dcolumn}   
\usepackage{bm}        
\usepackage{amssymb}   

\begin{document}



\title{Study of baryon acoustic oscillations with SDSS DR12 data
and measurement of $\Omega_\textrm{DE}(a)$}
\author{B.~Hoeneisen} \affiliation{Universidad San Francisco de Quito, Quito, Ecuador}

\date{\today}

\begin{abstract}
\noindent
We define Baryon Acoustic Oscillation (BAO) distances
$\hat{d}_\alpha(z, z_c)$, $\hat{d}_z(z, z_c)$, and $\hat{d}_/(z, z_c)$ that do not
depend on cosmological parameters. These BAO distances
are measured as a function of redshift $z$ with the
Sloan Digital Sky Survey (SDSS) data release DR12. 
From these BAO distances alone, or together 
with the correlation angle $\theta_\textrm{MC}$
of the Cosmic Microwave Background (CMB), we constrain
the cosmological parameters in several scenarios.
We find $4.3 \sigma$ tension between the BAO plus $\theta_\textrm{MC}$
data and a cosmology with flat space and constant dark energy 
density $\Omega_\textrm{DE}(a)$.
Releasing one and/or the other of these constraints obtains 
agreement with the data.
We measure $\Omega_\textrm{DE}(a)$ as a function of $a$.
\end{abstract}

\pacs{}
\maketitle

\section{Introduction}
A point-like peak in the primordial density of the universe results,
well after recombination and decoupling, in two spherical shells of
overdensity: one of radius $\approx 150$ Mpc, and one of
radius $\approx	18$ Mpc	\cite{Eisenstein, BAO1, BAO2}.
(All distances in this article are ``co-moving", i.e. are
referred to the present time $t_0$.) 
The ``acoustic length scale" $r'_S \approx 150$ Mpc
is approximately the distance that sound waves 
of the tightly coupled plasma of photons, electrons, protons, and helium nuclei
traveled from the time of the Big Bang until
the electrons recombined with the protons and helium nuclei
to form	neutral	atoms, and the photons decoupled.   
The inner spherical shell of $\approx 18$ Mpc becomes
re-processed by	the hierarchical formation of galaxies \cite{BH},
while the radius of the outer shell is unprocessed to better than 1\% \cite{BAO2}
(or even 0.1\% with corrections \cite{BAO2})
and therefore is an excellent standard ruler to	measure	the
expansion of the universe as a function	of redshift $z$.
Histograms of galaxy-galaxy distances show an excess of	galaxy pairs
with distances in the approximate range	$150 - 18$ to $150 + 18$ Mpc.
This ``Baryon Acoustic Oscillation" (BAO) signal has a signal-to-background
ratio of the order of 0.1\% to 1\%. Measurements of these BAO signals  
are by now well	established (see \cite{BAO1, BAO2} for extensive 
lists of early publications).

In this article we present studies of BAO
with Sloan Digital Sky Survey (SDSS) data release DR12 \cite{DR12}.

\section{The homogeneous universe}
\label{homo}
To establish the notation, let us recall the
equations that describe the metric and evolution of a
homogeneous universe in the General Theory of Relativity: \cite{PDG}
\begin{eqnarray}
ds^2 & = & dt^2 - R^2(t) \left[ \frac{dr^2}{1 - kr^2} + 
r^2(d\theta^2 + \sin^2{\theta} d\phi^2) \right], \nonumber \\
\label{s} \\
\frac{H}{H_0} & \equiv & E(a) =
\sqrt{\frac{\Omega_\textrm{m}}{a^3} + \frac{\Omega_{\textrm{r}}}{a^4} 
+ \Omega_\textrm{DE}(a) + \frac{\Omega_k}{a^2}}, \label{E}
\end{eqnarray}
where the expansion and Hubble parameters are defined as
\begin{equation}
a(t) \equiv \frac{R(t)}{R(t_0)}, \qquad
H(t) \equiv \frac{da/dt}{a}.
\end{equation}
We rescale $r$ such that the curvature constant $k$ adopts one of three
values: $k = 0, 1$ or $-1$ for a spatially flat, closed or
open universe, respectively.
Today $t = t_0$, $a(t_0) \equiv a_0 \equiv 1$ 
and $H(t_0) \equiv H_0$. 
The ``red shift" $z$ is related to the expansion parameter:
\begin{equation}
a = \frac{1}{1 + z} \equiv \frac{f_0}{f}
\end{equation}
where $f$ and $f_0$ are the frequencies of emission and reception of
a ray of light as measured by comoving observers.
Note that along a light ray $f a$ is constant.
$\Omega_\textrm{m}/a^3$ and $\Omega_\textrm{r}/a^4$ are, respectively, the densities
of non-relativistic matter and ultra-relativistic radiation measured
in units of the ``critical density"
$\rho_c \equiv 3H^2_0/(8 \pi G_N)$.
$\Omega_k \equiv -k/\left[R(t_0) H_0\right]^2$.
In the General Theory of Relativity $\Omega_\textrm{DE} = \Lambda/(3 H_0^2)$,
where $\Lambda$ is the ``cosmological constant". To test the theory
we let $\Omega_\textrm{DE}(a) \equiv \Omega_\textrm{DE} f(a)$ (with $f(1) = 1$) 
be a function of $a$ to be determined by observations.
Note that from Eq. (\ref{E}) at $t_0$ we obtain the identity
\begin{equation}
\Omega_\textrm{m} + \Omega_\textrm{r} + \Omega_\textrm{DE} + \Omega_k \equiv 1.
\end{equation}
It is observed that $\Omega_k \ll 1$ \cite{PDG} so we expand
expressions up to first order in $\Omega_k$.
At the late times of interest the density of radiation is
negligible with respect to the density of matter 
so we set $\Omega_\textrm{r} = 0$. 

We consider four scenarios:
\begin{enumerate}
\item
The observed acceleration of the expansion of the universe is due
to the cosmological constant, i.e. $f(a) = 1$.
\item
The observed acceleration of the expansion of the universe is due
to a gas of negative pressure with an equation of state $w \equiv p/\rho < 0$.
We allow the index $w$ to be a function of $a$ \cite{BAO1, Linder}:
$w(a) = w_0 + w_a (1 - a)$. While this gas dominates $E(a)$ the
equation \cite{PDG}
\begin{equation}
\frac{d \rho}{dt} = -3 \frac{da/dt}{a} ( \rho + p )
\end{equation}
can be integrated with the result \cite{BAO1, Linder}
\begin{equation}
f(a) = a^{-3(1 + w_0 + w_a)} \exp{\left\{ -3w_a (1 - a) \right\}}.
\end{equation}
If $w_0 = -1$ and $w_a = 0$ we obtain $f(a) = 1$ as in the General Theory
of Relativity. 
\item
Same as Scenario 2 with $w(a)$ constant, i.e. $w_a = 0$.
\item
We measure the dependence of $\Omega_\textrm{DE}(a)$ 
on $a$ in the linear approximation
\begin{equation}
f(a) = 1 + w_1 (1 - a).
\end{equation}
\end{enumerate}
Note that BAO measurements can constrain $\Omega_\textrm{DE}(a)$ 
for $0.5 \lesssim a \le 1$ where
$\Omega_\textrm{DE}(a)$ contributes significantly to $E(a)$.

\section{BAO observables}

Problem: An astronomer observes two galaxies at the same redshift $z$
separated by an angle $\alpha \lesssim 1$. What is the distance $d'_\alpha$ between
these two galaxies today? Answer to sufficient accuracy: 
\begin{eqnarray}
d'_\alpha & \equiv & \frac{c}{H_0} d_\alpha =
\frac{c}{H_0} 2 d_A(z) \sin{\left( \frac{\alpha}{2} \right)}, \nonumber  \\
d_A(z) & = & \chi(z) \left( 1 + \frac{1}{6} \Omega_k \chi^2(z) \right), \nonumber \\
\chi(z) & \equiv & \int_0^z{\frac{dz'}{E(z')}}.
\label{dalpha}
\end{eqnarray}
Here we have written the speed of light $c$ explicitly.
The function $\chi(z)$ is presented 
in Table \ref{chi} for some bench mark universes.
($\chi(z)$ is not to be confused with the $\chi^2$ of fits.)

\begin{table}
\caption{\label{chi}
Function $\chi(z)$ for several values of $z$, 
$\Omega_\textrm{m}$ and $\Omega_\textrm{DE}$. 
$\Omega_k = 1 - \Omega_\textrm{m} - \Omega_\textrm{DE}$.
Measurements at the ``Lyman alpha forest" correspond to $z \approx 2.4$.
Decoupling of photons occurs at $z = 1090.2 \pm 0.7$ \cite{PDG}.
All columns have $w_0 = -1$, $w_a = 0$, except the row (*) which
has $w_0 = -0.9$, $w_a = 0$.
An approximation to $\chi(z)$ good to
an accuracy of $\pm 1\%$ for $z$ in the range 0 to 1 is
$\chi(z) = z \exp(-z/z_c)$ with $z_c$ given in the next-to-last row.
$r_S$ is calculated from Eq. (\ref{thetaMC_rs}) with $\theta_\textrm{MC}$ given by
(\ref{tMC}).
}
\begin{ruledtabular}
\begin{tabular}{c|ccccc}
$\Omega_\textrm{m}$          & 0.25 & 0.30 & 0.35 & 0.25 & 0.30 (*) \\
$\Omega_\textrm{DE}$       & 0.75 & 0.70 & 0.65 & 0.70 & 0.70 \\
$\Omega_k$          & 0.00 & 0.00 & 0.00 & 0.05 & 0.00 \\
\hline
$z = 0.2$    & 0.192 & 0.191 & 0.189 & 0.191 & 0.189 \\
$z = 0.4$    & 0.368 & 0.362 & 0.357 & 0.364 & 0.357 \\
$z = 0.6$    & 0.526 & 0.515 & 0.505 & 0.520 & 0.506 \\
$z = 0.8$    & 0.668 & 0.651 & 0.635 & 0.659 & 0.638 \\
$z = 1.0$    & 0.796 & 0.771 & 0.750 & 0.783 & 0.756 \\
$z = 2.4$    & 1.400 & 1.333 & 1.276 & 1.371 & 1.306 \\
$z = 3.0$    & 1.564 & 1.484 & 1.417 & 1.532 & 1.456 \\
$z = 1090.2$ & 3.416 & 3.178 & 2.988 & 3.368 & 3.147 \\
\hline
$z_c$      & 4.33 & 3.84 & 3.48 & 4.07 & 3.58 \\
$r_S \times 100$     & 3.557 & 3.309 & 3.112 & 3.839 & 3.277 \\
\end{tabular}
\end{ruledtabular}
\end{table}

\begin{table}
\caption{\label{E_table}
Function $1/E(z)$ for several values of $z$, 
$\Omega_\textrm{m}$ and $\Omega_\textrm{DE}$. 
$\Omega_k = 1 - \Omega_\textrm{m} - \Omega_\textrm{DE}$.
All columns have $w_0 = -1$, $w_a = 0$, except the row (*) which
has $w_0 = -0.9$, $w_a = 0$.
An approximation to $1/E(z)$ good to
an accuracy of $\pm 1\%$ for $z$ in the range 0 to 1 is
$1/E(z) = (1 - z/z_c) \exp(-z/z_c)$ with $z_c$ given in the last row.
}
\begin{ruledtabular}
\begin{tabular}{c|ccccc}
$\Omega_\textrm{m}$          & 0.25 & 0.30 & 0.35 & 0.25 & 0.30 (*) \\
$\Omega_\textrm{DE}$       & 0.75 & 0.70 & 0.65 & 0.70 & 0.70 \\
$\Omega_k$          & 0.00 & 0.00 & 0.00 & 0.05 & 0.00 \\
\hline
$z = 0.2$    & 0.920 & 0.906 & 0.893 & 0.911 & 0.892 \\
$z = 0.4$    & 0.834 & 0.810 & 0.788 & 0.821 & 0.791 \\
$z = 0.6$    & 0.751 & 0.720 & 0.693 & 0.735 & 0.701 \\
$z = 0.8$    & 0.673 & 0.639 & 0.610 & 0.657 & 0.622 \\
$z = 1.0$    & 0.603 & 0.568 & 0.538 & 0.587 & 0.554 \\
\hline
$z_c$      & 4.33 & 3.84 & 3.48 & 4.07 & 3.58 \\
\end{tabular}
\end{ruledtabular}
\end{table}

Problem: An astronomer observes two galaxies at redshifts $z_1$
and $z_2$ with a negligible angle of separation
$\alpha$. What is the distance $d'_z$ between
these two galaxies today? Answer: 
\begin{eqnarray}
d'_z & \equiv & \frac{c}{H_0} d_z = \frac{c}{H_0} \left[ \chi(z_1) - \chi(z_2) \right].
\end{eqnarray}
For small separation $z_1 - z_2 = \Delta z$,
$d_z \approx \Delta z / E(z)$.
The function $1/E(z)$ is presented in Table \ref{E} for some bench mark universes.

Problem: An astronomer observes two galaxies at redshifts $z_1$ 
and $z_2$ separated by an angle $\alpha \lesssim 1$. 
What is the distance $d'$ between
these two galaxies today? Answer: $d' \equiv (c / H_0) d$, where, 
to sufficient accuracy,
\begin{eqnarray}
d & = & \sqrt{ d^2_\alpha + d^2_z }, \nonumber \\
d_\alpha & = & 2 \sin{\left( \frac{\alpha}{2} \right)} \sqrt{\chi(z_1) \chi(z_2)} 
\left[ 1 + \frac{1}{6} \Omega_k \chi(z_1) \chi(z_2) \right], \nonumber \\
d_z & = & \chi(z_1) - \chi(z_2). 
\label{dprime}
\end{eqnarray}
The distances $d$, $d_\alpha$ and $d_z$ are dimensionless: 
they are expressed in units of $c/H_0$. This way we decouple
BAO observables from the uncertainty of $H_0$.

Equations (\ref{dalpha}) can be applied to the cosmic microwave background (CMB).
It is observed that fluctuations in the CMB have a correlation angle \cite{PDG}
\begin{equation}
\theta_\textrm{MC} = 0.010413 \pm 0.000006
\label{tMC}
\end{equation}
(we have chosen a measurement with no input from BAO).
The extreme precision with which $\theta_\textrm{MC}$ is measured 
makes it one of the primary parameters of cosmology.
The corresponding distance today is
 \begin{equation}
r'_S = \frac{c}{H_0} 2 d_A(z_\textrm{dec}) 
\sin{ \left( \frac{1}{2} \theta_\textrm{MC} \right) },
\label{rS}
\end{equation}
and depends on the cosmological parameters.
For the six-parameter $\Lambda$CDM
cosmology fit to Plank CMB data \cite{PDG}, 
\begin{equation}
r'_S = 147.5 \pm 0.6 \textrm{ Mpc}.
\end{equation}
From Eqs. (\ref{dalpha}) with $H_0 \equiv 100$ $h$ km s$^{-1}$ Mpc$^{-1}$, 
$h = 0.673 \pm 0.012$ \cite{PDG}, and $\chi(z_\textrm{dec}) \approx 3.178$
from Table \ref{chi}, we obtain $r'_S(z_\textrm{dec}) \approx 147.4$ Mpc
and $r_S \equiv (H_0/c) r'_S(z_\textrm{dec}) \approx 0.0331$.
We expect an excess of galaxy pairs with a present day separation 
$r'_S$.

We find the following approximations to
$\chi(z)$ and $1/E(z)$ valid in the range $0 \le z < 1$
with precision $\pm 1\%$:
\begin{equation}
\chi(z) \approx z \exp{\left( - \frac{z}{z_c} \right)}, \qquad 
\frac{1}{E(z)} \approx \left( 1 - \frac{z}{z_c} \right) \exp{\left( - \frac{z}{z_c} \right)}.
\label{chi_approx}
\end{equation}
The values of $z_c$ for some bench-mark universes are given
in Tables \ref{chi} or \ref{E_table}.

Our strategy is as follows:
We consider galaxies with red shift in a given range 
$z_{\textrm{min}} < z < z_{\textrm{max}}$.
For each galaxy pair we calculate $d_\alpha(z, z_c)$, $d_z(z, z_c)$ and $d(z, z_c)$
with Eqs. (\ref{dprime}) \textit{with the approximation} (\ref{chi_approx}), 
\textit{and with} $\Omega_k = 0$,
and fill one of three histograms of $d(z, z_c)$ (with weights to
be discussed later)
depending on the ratio $d_z(z, z_c) / d_\alpha(z, z_c)$:
\begin{itemize}
\item
If $d_z(z, z_c) / d_\alpha(z, z_c) < 1/3$ fill a histogram of $d(z, z_c)$ that obtains
a BAO signal centered at $\hat{d}_\alpha(z, z_c)$. For this histogram,
$d^2_z(z, z_c)$ is a small correction relative to $d^2_\alpha(z, z_c)$ that is calculated
with sufficient accuracy with the approximation (\ref{chi_approx})
and $\Omega_k = 0$.
\item
If $d_\alpha(z, z_c) / d_z(z, z_c) < 1/3$ fill a second histogram of $d(z, z_c)$
that obtains a BAO signal centered at $\hat{d}_z(z, z_c)$. For this histogram,
$d^2_\alpha(z, z_c)$ is a small correction relative to $d^2_z(z, z_c)$ that is calculated
with sufficient accuracy with the approximation (\ref{chi_approx})
and $\Omega_k = 0$.
\item
Else, fill a third histogram of $d(z, z_c)$ that obtains a 
BAO signal centered at $\hat{d}_/(z, z_c)$. 
\end{itemize}
BAO observables $\hat{d}_\alpha(z, z_c)$, $\hat{d}_z(z, z_c)$, and $\hat{d}_/(z, z_c)$
were chosen because (i) they are dimensionless,
(ii) they do not depend on any
cosmological parameter, and (iii) are
almost independent of $z$ 
(for an optimized value of $r_c \approx 3.79$) so that
a large bin $z_\textrm{max} - z_\textrm{min}$ may be analyzed.

The BAO distance $d_\textrm{BAO}$ is obtained from the BAO 
observables $\hat{d}_\alpha(z, z_c)$, $\hat{d}_z(z, z_c)$, or $\hat{d}_/(z, z_c)$
as follows:
\begin{eqnarray}
d_\textrm{BAO} & = & \hat{d}_\alpha (z, z_c) \frac{\chi(z) \left[ 1 + \frac{1}{6} \Omega_k \chi^2(z) \right]}
 {z \exp{(-z/z_c)}},  \label{da_rs} \\
d_\textrm{BAO} & = & \hat{d}_z (z, z_c) \frac{1}{(1 - z/z_c) \exp{(-z/z_c)} E(z)}, \label{dz_rs} \\
d_\textrm{BAO} & = & \hat{d}_/(z, z_c) \left(\frac{\chi(z) \left[ 1 + \frac{1}{6} \Omega_k \chi^2(z) \right]}
 {z \exp{(-z/z_c)}}\right)^{n/3} \nonumber \\
& & \times \left(\frac{1}{(1 - z/z_c) \exp{(-z/z_c)} E(z)}\right)^{1-n/3}. \label{d_rs}
\end{eqnarray}
The dimensionless correlation distance $r_S$ is obtained from
$\theta_\textrm{MC}$ as follows:
\begin{eqnarray}
r_S & = & 2 \sin{\left( \frac{1}{2} \theta_\textrm{MC} \right)} \chi(z_\textrm{dec}) \nonumber \\
& & \times \left[ 1 + \frac{1}{6} \Omega_k \chi^2(z_\textrm{dec}) \right] \label{thetaMC_rs}.
\label{rS2}
\end{eqnarray}
For $\chi(z_\textrm{dec})$ we do not neglect $\Omega_\textrm{r}$ of photons,
or three generations of massless Dirac neutrinos, i.e. $N_\textrm{eq} = 3.36$.
We set $r_S = d_\textrm{BAO}$.

A numerical analysis obtains $n = 1.70$ for $z = 0.2$, dropping to
$n = 1.66$ for $z = 0.8$ (in agreement with the method introduced in \cite{Eisenstein}
that obtains $n = 2$ when $d$ covers all angles).
Note that $d_\textrm{BAO} = r_S$, but we use different notations
because $r_S$ given by Eq. (\ref{rS2})
depends on cosmological parameters while $d_\textrm{BAO}$ does not, and
$d_\textrm{BAO}$ and $r_S$
have different systematic uncertainties
and may require different corrections.
The redshifts $z$ in Eqs. (\ref{da_rs}), (\ref{dz_rs}) and (\ref{d_rs})
correspond to the weighted mean of $z$ in the range $z_{\textrm{min}}$ to $z_{\textrm{max}}$.
The fractions in Eqs. (\ref{da_rs}), (\ref{dz_rs}) and (\ref{d_rs})
are very close to 1 for $z_c \approx 3.79$.
Note that the limits of $\hat{d}_\alpha (z, z_c)$ or $\hat{d}_z (z, z_c)$ or 
$\hat{d}_/(z, z_c)$ as $z \rightarrow 0$ are all equal to $d_\textrm{BAO}$.

\section{Galaxy selection and data analysis}

We obtain the following data from the SDSS DR12 catalog \cite{DR12} for all objects
identified as galaxies that pass quality selection flags:
right ascension ra, declination dec, redshift $z$, redshift uncertainty $zErr$,
and the absolute values of magnitudes $g$ and $r$. There are 991504 such galaxies.
We require a good measurement of redshift, i.e. $zErr < 0.001$, leaving
987933 galaxies. To have well defined edge effects, we limit the present
study to galaxies with right ascension in the range $110^0$ to $260^0$,
and declination in the range $0^0$ to $70^0$. This selection leaves 
832507 galaxies (G). 

We renormalize the magnitudes $g$ and $r$ to a common redshift
$z = 0.35$ ($g_{35}$ and $r_{35}$ respectively) 
and calculate their absolute luminosities in the $r$ band relative
to the absolute luminosity corresponding to $r_{35} = 19.0$ ($F$).
We define ``luminous galaxy" (LG), e.g. $r_{35} < 19.8$, 
``luminous red galaxy" (LRG), e.g. $r_{35} < 19.5$
and $g_{35} - r_{35} > 1.65$,
``clusters" (C), and ``large clusters" (LC).
Clusters C are based on a cluster finding algorithm that uses LG's as seeds.
Large clusters LC are based on a cluster finding algorithm that obtains
averages (with weights $F$) of ra, dec and $z$ of galaxies in cubes of
size $(0.017 c/H_0)^3$ and then selects cubes with a total absolute luminosity
greater than a minimum such that the cube occupancy is less than 0.5.
We define ``field galaxy" (FG) as a galaxy with $d > 0.003$ from any C.
We define ``cluster galaxy" (CG) as a galaxy with $d < 0.003$ from at
least one C.

We define a ``run" by specifying a range of redshifts 
$(z_\textrm{min}, z_\textrm{max})$, and a selection of 
``galaxies" (G, LG, FG, or CG) and ``centers"
(G, LG, LRG, C, or LC). We fill histograms
of center-galaxy distances and obtain the BAO distances 
$\hat{d}_\alpha(z, z_c)$, $\hat{d}_z(z, z_c)$, and $\hat{d}_/(z, z_c)$
by fitting these histograms. Histograms are filled with weights $(0.033/d)^2$
or $F_i F_j (0.033/d)^2$ where $F_i$ and $F_j$ are the absolute luminosities
$F$ of galaxies $i$ and centers $j$ respectively. 
We obtain histograms with $z_c = 3.79, 3.0$ and $5.0$.
We repeat the measurements with fine and coarse binnings of $z$,
and with overlapping bins of $z$. 
The reason for this large degree of redundancy is the
difficulty to discriminate the BAO signal from the background with its
statistical and cosmological fluctuations due to galaxy clustering. 

\begin{figure}
\begin{center}
\scalebox{0.45}
{\includegraphics{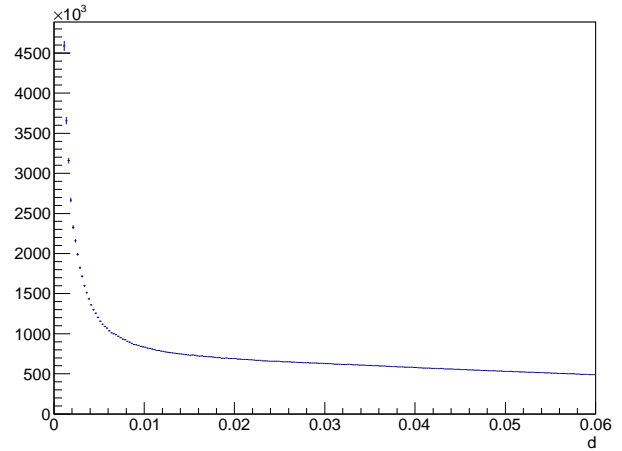}}
\caption{
Histogram of the galaxy-galaxy BAO distances $d(z, z_c)$
with weights $(0.033/d)^2$ for galaxy pairs
with $d_z(z, z_c) < d_\alpha(z, z_c)/3$  
corresponding to the entry with $z = 0.54$ 
in Table \ref{d0}. This histogram obtains 
$\hat{d}_\alpha(0.54, z_c)$ as shown in Fig. \ref{fig_d_fit}.
Note galaxy clustering at small $d$.
}
\label{fig_d}
\end{center}
\end{figure}

\begin{figure}
\begin{center}
\scalebox{0.45}
{\includegraphics{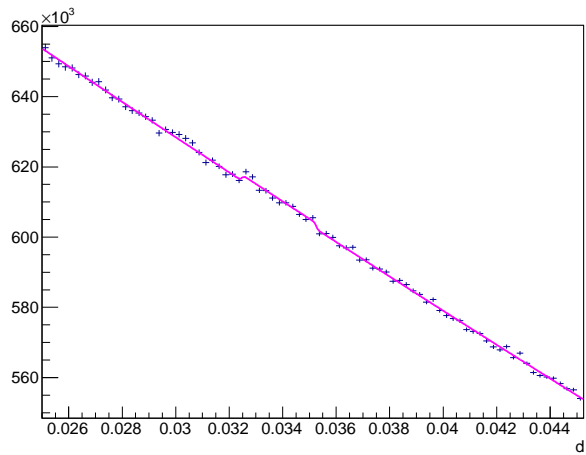}}
\caption{
Detail of Fig. \ref{fig_d} with fit that obtains 
$\hat{d}_\alpha(0.54, z_c)$. Note the fluctuations
in the background due to galaxy clustering.
}
\label{fig_d_fit}
\end{center}
\end{figure}

The fitting function is a second degree polynomial for the
background and, for the BAO signal, a step-up-step-down function
of the form
\begin{eqnarray*}
\frac{\exp{(x_<)}}{\exp{(x_<)} + \exp{(-x_<)}} - 
\frac{\exp{(x_>)}}{\exp{(x_>)} + \exp{(-x_>)}}
\end{eqnarray*}
where
\begin{eqnarray*}
x_< = \frac{d - \hat{d} + \Delta d}{\sigma}, \qquad 
x_> = \frac{d - \hat{d} - \Delta d}{\sigma}. 
\end{eqnarray*}
An example of a BAO distance
histogram is presented in Fig. \ref{fig_d}.
A close-up of the fit to the BAO signal in Fig. \ref{fig_d}
is presented in Fig. \ref{fig_d_fit}.
The most prominent features of the BAO signal are its
lower edge at $\hat{d} - \Delta d$ and its upper edge at
$\hat{d} + \Delta d$, see Fig. \ref{fig_d_fit}. 

With few exceptions, we only accept runs with good fits for
all three BAO distances $\hat{d}_\alpha(z, z_c)$,
$\hat{d}_z(z, z_c)$, and $\hat{d}_/(z, z_c)$ with
consistent relative amplitudes and half-widths $\Delta d$
of the signal.

Final results of these BAO distance measurements are presented
in Table \ref{d0}. These 18 BAO distances are independent,
do not depend on cosmological parameters, and
are the main result
of the present analysis. As cross-checks, and to estimate the
systematic uncertainties, we present additional BAO distance 
measurements in Table \ref{d1}.

\section{Systematic uncertainties}

The backgrounds of BAO distance histograms have fluctuations
due to the clustering of galaxies \cite{BH} 
as seen in Fig. \ref{fig_d_fit}. These fluctuations are
the dominant source of systematic uncertainties of the
BAO distance measurements. These systematic effects are
independent for each entry in Table \ref{d0}.
We estimate the systematic uncertainties directly from the data 
by calculating the standard deviations of measurements 
presented in Tables \ref{d0} and \ref{d1} with similar
$z$ separately for $\hat{d}_\alpha(z, z_c)$,
$\hat{d}_z(z, z_c)$, and $\hat{d}_/(z, z_c)$.
These measurements have different galaxy and center
selections and different bins $(z_\textrm{min}, z_\textrm{max})$
so the fluctuations of the backgrounds of
the BAO distance histograms are different.
The resulting standard deviations are approximately $0.00060$ for
$\hat{d}_\alpha(z, z_c)$,
$\hat{d}_z(z, z_c)$, and $\hat{d}_/(z, z_c)$ independently of $z$, so we 
assign an independent systematic uncertainty 
$\pm 0.00060$ to each entry in Tables \ref{d0} and \ref{d1}
to be summed in quadrature with the statistical 
uncertainties obtained from the fits.
The histograms of BAO distances have a bin size 
$0.00025$ that is much smaller than the systematic uncertainty.

Here are two ideas left for future studies:
(i) The systematic uncertainty $\pm 0.00060$ is about three times
larger than the statistical uncertainties from the fits.
Therefore it should be possible to improve the selection
of galaxies and centers to reduce the systematic uncertainty
at the expense of increasing the statistical uncertainties.
(ii) For some bins in $z$ it was possible to fit several runs,
e.g. G-G, LG-LG and G-C, which are partially independent, 
see Tables \ref{d0} and \ref{d1}. 
Averaging such measurements may be considered.

\begin{table*}
\caption{\label{d0}
Independent measured BAO distances $\hat{d}_\alpha(z, z_c)$, 
$\hat{d}_z(z, z_c)$, and $\hat{d}_/(z, z_c)$ in units of $c/H_0$ 
with $z_c = 3.79$ (see text) from SDSS DR12 galaxies with
right ascension $110^0$ to $260^0$ and declination $0^0$ to $70^0$.
Uncertainties are statistical from the fits to the BAO signal. 
Each BAO distance has an independent systematic uncertainty $\pm 0.00060$.
No corrections have been applied.
}
\begin{ruledtabular}
\begin{tabular}{c|ccccc|ccc}
$z$ & $z_\textrm{min}$ & $z_\textrm{max}$ & galaxies & centers & type &
$\hat{d}_\alpha(z, z_c) \times 100$ & $\hat{d}_z(z, z_c) \times 100$ & $\hat{d}_/(z, z_c) \times 100$ \\
\hline
$0.10$ & 0.0 & 0.2 & 297220& 5770 & G-C & $3.338 \pm 0.003$ & $3.354 \pm 0.006$ & $3.361 \pm 0.017$ \\
$0.25$ & 0.2 & 0.3 & 56545 & 3552 & G-LG & $3.454 \pm 0.016$ & $3.301 \pm 0.008$ & $3.432 \pm 0.011$ \\
$0.35$ & 0.3 & 0.4 & 78634 & 32095 & G-LG & $3.318 \pm 0.006$ & $3.276 \pm 0.013$ & $3.410 \pm 0.009$ \\
$0.46$ & 0.4 & 0.5 & 124225 & 124225 & G-G & $3.535 \pm 0.013$ & $3.494 \pm 0.017$ & $3.442 \pm 0.036$ \\
$0.54$ & 0.5 & 0.6 & 168766 & 168766 & G-G & $3.385 \pm 0.007$ & $3.450 \pm 0.019$ & $3.547 \pm 0.014$ \\
$0.67$ & 0.6 & 0.9 & 95206 & 2617 & G-C & $3.485 \pm 0.019$ & $3.471 \pm 0.027$ & $3.364 \pm 0.070$ \\
\end{tabular}
\end{ruledtabular}
\end{table*}

\begin{table*}
\caption{\label{d1}
Additional measured BAO distances $\hat{d}_\alpha(z, z_c)$, 
$\hat{d}_z(z, z_c)$, and $\hat{d}_/(z, z_c)$ in units of $c/H_0$ 
with $z_c = 3.79$ (see text) from SDSS DR12 galaxies with
right ascension $110^0$ to $260^0$ and declination $0^0$ to $70^0$.
Uncertainties are statistical from the fits to the BAO signal. 
Each BAO distance has an independent systematic uncertainty $\pm 0.00060$.
No corrections have been applied.
}
\begin{ruledtabular}
\begin{tabular}{c|ccccc|ccc}
$z$ & $z_\textrm{min}$ & $z_\textrm{max}$ & galaxies & centers & type &
$\hat{d}_\alpha(z, z_c) \times 100$ & $\hat{d}_z(z, z_c) \times 100$ & $\hat{d}_/(z, z_c) \times 100$ \\
\hline
$0.13$ & 0.1 & 0.2 & 175841 & 175841 & G-G & $$ & $3.450 \pm 0.012$ &  \\
$0.25$ & 0.2 & 0.3 & 14699 & 4026 & FG-C & $3.431 \pm 0.015$ &  &  \\
$0.25$ & 0.2 & 0.3 & 41846 & 4026 & CG-C & & $3.335 \pm 0.010$ & $3.389 \pm 0.019$ \\ 
$0.32$ & 0.2 & 0.4 & 135179 & 16446 & G-C & $3.427 \pm 0.014$ & $3.413 \pm 0.019$ & $3.457 \pm 0.011$ \\
$0.35$ & 0.3 & 0.4 & 26130 & 5112 & FG-C & & & $3.345 \pm 0.018$ \\
$0.38$ & 0.2 & 0.5 & 237651 & 25318 & G-C & $3.388 \pm 0.058$ & $3.416 \pm 0.016$ & $3.404 \pm 0.011$ \\
$0.38$ & 0.2 & 0.5 & 259404 & 259404 & G-G & $3.588 \pm 0.014$ & $3.329 \pm 0.008$ & $3.382 \pm 0.009$ \\
$0.38$ & 0.2 & 0.5 & 259404 & 28515 & G-C & & $3.413 \pm 0.017$ & $3.431 \pm 0.008$ \\ 
$0.41$ & 0.3 & 0.5 & 202859 & 45033 & G-LG & $3.398 \pm 0.060$ & $3.318 \pm 0.036$ & $3.387 \pm 0.009$ \\
$0.46$ & 0.4 & 0.5 & 124225 & 12089 & G-C & $3.501 \pm 0.012$ & $3.445 \pm 0.033$ & $3.435 \pm 0.015$ \\
$0.46$ & 0.4 & 0.5 & 68361 & 4945 & FG-C & $3.337 \pm 0.021$ & & $3.453 \pm 0.015$ \\
$0.54$ & 0.5 & 0.6 & 168766 & 11671 & G-C & $3.429 \pm 0.015$ & $3.637 \pm 0.026$ & $3.627 \pm 0.016$ \\
$0.54$ & 0.5 & 0.6 & 168766 & 5553 & G-LC & $3.373 \pm 0.018$ & $3.485 \pm 0.010$ & $3.423 \pm 0.012$ \\
$0.54$ & 0.5 & 0.6 & 168766 & 1608 & G-C & $3.545 \pm 0.034$ & $3.568 \pm 0.013$ & $3.578 \pm 0.054$ \\ 
$0.54$ & 0.5 & 0.6 & 49102 & 4180 & CG-C & $3.453 \pm 0.013$ & $3.404 \pm 0.171$ &  \\ 
$0.59$ & 0.5 & 0.9 & 263973 & 1959 & G-C & $3.557 \pm 0.028$ & $3.531 \pm 0.031$ & $3.576 \pm 0.026$ \\
$0.59$ & 0.5 & 0.9 & 120927 & 120927 & LG-LG & $3.414 \pm 0.071$ & $3.586 \pm 0.083$ & $3.530 \pm 0.017$ \\
$0.59$ & 0.5 & 0.9 & 263973 & 263973 & G-G & $3.380 \pm 0.008$ & $3.560 \pm 0.013$ & $3.576 \pm 0.011$ \\
$0.67$ & 0.6 & 0.9 & 95206  & 95206 & G-G & $3.397 \pm 0.019$ & $$ & $3.492 \pm 0.009$ \\
$0.67$ & 0.6 & 0.9 & 51310  & 51310 & LG-LG & $3.469 \pm 0.017$ & $3.586 \pm 0.018$ & $3.459 \pm 0.014$ \\
$0.67$ & 0.6 & 0.9 & 95206  & 6741 & G-C & $3.505 \pm 0.020$ & $$ & $3.459 \pm 0.014$ \\
$0.67$ & 0.6 & 0.9 & 109488 & 109488 & G-G & $3.391 \pm 0.121$ & $3.549 \pm 0.014$ & $3.397 \pm 0.025$ \\
\end{tabular}
\end{ruledtabular}
\end{table*}

\section{Corrections}

Let us consider	corrections to the BAO distances 
due to peculiar	velocities and peculiar displacements
of galaxies towards their centers.
A relative peculiar velocity $v_p$ towards the center
causes a reduction of the BAO distances
$\hat{d}_z(z, z_c)$, $\hat{d}_/(z, z_c)$, and $\hat{d}_\alpha(z, z_c)$
of order $v_p/c$. In addition, the Doppler shift
produces an apparent shortening of $\hat{d}_z(z, z_c)$
by $v_p/c$, and somewhat less for $\hat{d}_/(z, z_c)$.

From the studies in Ref.~\cite{Scoccimarro} we add the
following corrections to the measured BAO distances
presented in Tables \ref{d0} and \ref{d1}:
\begin{equation}
\Delta d_z = 0.0012, \Delta d_/ = 0.75 \Delta d_z, \Delta d_\alpha = 0.5 \Delta d_z
\label{Dd}
\end{equation}
respectively to $\hat{d}_z(z, z_c)$, $\hat{d}_/(z, z_c)$, 
and $\hat{d}_\alpha(z, z_c)$.
These corrections depend on $z$ and on the mass of the halos
so $\Delta d_z$ lies in the approximate range 0.0006 to 
0.0024~\cite{Scoccimarro}.

We would like to obtain the peculiar motion corrections 
directly from the data. We exploit the fact that the corrections
$\Delta d_z$ and $\Delta d_\alpha \approx 0.5 \Delta d_z$ \cite{Scoccimarro}
are different so the correct $\Delta d_z$ should minimize
the $\chi^2$ of fits. In Table \ref{BAO_Ddz} we present fits to 
the cosmological parameters $\Omega_k$, $\Omega_\textrm{DE} - 0.5 \Omega_k$
and $d_\textrm{BAO}$
that minimize the $\chi^2$ with
18 terms corresponding to the 18 entries in Table \ref{d0}
for $\Delta d_z = 0.0000$, $0.0012$, $0.0024$, and $0.0036$.
The fits correspond to Scenario 4 with free $\Omega_k$. We present 
$\Omega_\textrm{DE} + 0.5 \Omega_k$ which has a smaller
uncertainty than $\Omega_\textrm{DE}$. We observe
that the $\chi^2$ of the fits is minimized at $\Delta d_z = 0.0012 \pm 0.0013$.
The same result $\Delta d_z = 0.0012 \pm 0.0013$ is obtained for Scenario 1
with free $\Omega_k$. 
Minimizing the $\chi^2$ with 19 terms corresponding to
the 18 BAO measurements plus the measurement of $\theta_\textrm{MC}$
in Scenario 4 with free $\Omega_k$
results in a minimum $\chi^2$ at $\Delta d_z = 0.0018 \pm 0.0008$.

\begin{table*}
\caption{\label{BAO_Ddz}
Cosmological parameters obtained from the 18 BAO measurements in Table \ref{d0}
in Scenario 4, i.e. 
$\Omega_\textrm{DE}(a) = \Omega_\textrm{DE} \left[1 + w_1 (1 - a)\right]$,
for several peculiar motion corrections $\Delta d_z$, see Eq. (\ref{Dd}).
The minimum $\chi^2$ is obtained at $\Delta d_z = 0.0012 \pm 0.0013$.
}
\begin{ruledtabular}
\begin{tabular}{c|ccccc} 
   & Scenario 4 & Scenario 4 & Scenario 4 & Scenario 4 \\
$\Delta d_z$   & 0.0000 & 0.0012 & 0.0024 & 0.0036 \\
\hline
$\Omega_k$ & $-0.560 \pm 0.277$ & $-0.235 \pm 0.283$ & $0.096 \pm 0.288$ & $0.433 \pm 0.291$ \\
$\Omega_\textrm{DE} + 0.5 \Omega_k$ & $0.693 \pm 0.066$ & $0.656 \pm 0.065$ & $0.618 \pm 0.064$ 
  & $0.581 \pm 0.063$ \\
$w_1$ & $0.254 \pm 0.612$ & $0.132 \pm 0.783$ & $-0.085 \pm 1.096$ & $-0.558 \pm 1.851$ \\
$d_\textrm{BAO} \times 100$      & $3.38 \pm 0.06$ & $3.45 \pm 0.07$ & $3.51 \pm 0.07$ 
  & $3.57 \pm 0.07$ \\
$\chi^2/$d.f.                    & $19.4/14$ & $18.6/14$ & $19.3/14$ & $21.3/14$ \\
\end{tabular}
\end{ruledtabular}
\end{table*}

The distribution of the galaxies peculiar velocity component 
in one direction peaks at zero and drops to about 10\%
at $v_p \approx \pm 1000$ km/s \cite{BH} corresponding to $v_p/c = 0.0033$.
Most of these galaxies at large $v_p$
are in clusters. For relative peculiar velocities
of galaxies towards centers at the large BAO distance we expect
$v_p/c \ll 0.0033$.

In the limit $z \rightarrow 0$ we obtain the corrected BAO distances
$\hat{d}_z(0, z_c) = \hat{d}_\alpha(0, z_c)$.
If from \cite{Scoccimarro} we take approximately $\Delta d_\alpha \approx 0.5 \Delta d_z$,
we obtain $\Delta d_z \approx -0.0003 \pm 0.0017$ from the row $z = 0.1$ of Table \ref{d0}, and 
$\Delta d_z \approx 0.0031 \pm 0.0013$ from the row $z = 0.25$, with an average
$\Delta d_z \approx 0.0014 \pm 0.0011$.

In conclusion, we apply to the entries of Tables \ref{d0} and \ref{d1}
the corrections in Eqs. (\ref{Dd}) with $\Delta d_z = 0.0012 \pm 0.0012$,
which is coherent for all entries. In the next Section we present fits
for $\Delta d_z = 0.0012$ and $\Delta d_z = 0.0000$.

\begin{table*}
\caption{\label{BAO_fit}
Cosmological parameters obtained from the 18 BAO measurements in Table \ref{d0}
in several scenarios. Corrections for peculiar motions are given by
Eq. (\ref{Dd}) with the indicated $\Delta d_z$.
Scenario 1 has $\Omega_\textrm{DE}$ constant.
Scenario 2 has $w(a) = w_0 + w_a (1 - a)$.
Scenario 3 has $w = w_0$.
Scenario 4 has $\Omega_\textrm{DE}(a) = \Omega_\textrm{DE} \left[1 + w_1 (1 - a)\right]$.
}
\begin{ruledtabular}
\begin{tabular}{c|cccccc} 
   & Scenario 1 & Scenario 1 & Scenario 1 & Scenario 2 & Scenario 3 & Scenario 4 \\
$\Delta d_z$ & 0.0000 & 0.0012 & 0.0012 & 0.0012 & 0.0012 & 0.0012 \\
\hline
$\Omega_k$ & $0$ fixed  & $0$ fixed  & $-0.216 \pm 0.257$ & $-0.240 \pm 0.283$ & $-0.238 \pm 0.283$ 
  & $-0.235 \pm 0.283$ \\
$\Omega_\textrm{DE} + 0.5 \Omega_k$ & $0.660 \pm 0.022$ & $0.642 \pm 0.022$ & $0.646 \pm 0.022$ 
  & $0.830 \pm 0.797$ & $0.663 \pm 0.099$ & $0.656 \pm 0.065$ \\
$w_0$ & n.a. & n.a. & n.a. & $-0.821 \pm 0.729$ & $-0.948 \pm 0.270$ & n.a. \\
$w_a$ or $w_1$& n.a. & n.a. & n.a. & $0.754 \pm 1.496$ & n.a. & $0.132 \pm 0.783$ \\
$d_\textrm{BAO} \times 100$ & $3.38 \pm 0.03$ & $3.44 \pm 0.03$ & $3.45 \pm 0.03$ 
  & $3.45 \pm 0.07$ & $3.45 \pm 0.06$ & $3.45 \pm 0.07$ \\
$\chi^2/$d.f. & $23.4/16$ & $19.3/16$ & $18.6/15$ & $18.5/13$ & $18.6/14$ & $18.6/14$ \\
\end{tabular}
\end{ruledtabular}
\end{table*}

\begin{table*}
\caption{\label{BAO_thetaMC_fit_12}
Cosmological parameters obtained from the 18 BAO measurements in Table \ref{d0}
plus $\theta_\textrm{MC}$ from Eq. (\ref{tMC}) in several scenarios.
Corrections for peculiar motions are given by
Eq. (\ref{Dd}) with $\Delta d_z = 0.0012$.
Scenario 1 has $\Omega_\textrm{DE}$ constant.
Scenario 2 has $w(a) = w_0 + w_a (1 - a)$.
Scenario 3 has $w = w_0$.
Scenario 4 has $\Omega_\textrm{DE}(a) = \Omega_\textrm{DE} \left[1 + w_1 (1 - a)\right]$.
}
\begin{ruledtabular}
\begin{tabular}{c|cccccc}
   & Scenario 1 & Scenario 1 & Scenario 2 & Scenario 3 & Scenario 4 & Scenario 4 \\
$\Delta d_z$ & $0.0012$ & $0.0012$ & $0.0012$ & $0.0012$ & $0.0012$ & $0.0012$ \\
\hline
$\Omega_k$ & $0$ fixed  & $0.061 \pm 0.012$ & $0$ fixed & 
  $0$ fixed & $0$ fixed & $0.065 \pm 0.043$ \\
$\Omega_\textrm{DE} + 2 \Omega_k$ & $0.754 \pm 0.004$ & $0.733 \pm 0.005$ & $0.808 \pm 0.075$ 
  & $0.745 \pm 0.004$ & $0.732 \pm 0.006$ & $0.734 \pm 0.008$ \\
$w_0$ & n.a. & n.a. & $-0.795 \pm 0.092$ & $-0.730 \pm 0.046$ & n.a. & n.a. \\
$w_a$ or $w_1$ & n.a. & n.a. & $0.705 \pm 0.170$ & n.a. & $1.157 \pm 0.245$ & $-0.113 \pm 0.985$ \\
$d_\textrm{BAO} \times 100$ & $3.58 \pm 0.02$ & $3.44 \pm 0.03$ & $3.44 \pm 0.04$ 
  & $3.40 \pm 0.04$ & $3.37 \pm 0.04$ & $3.45 \pm 0.06$ \\
$\chi^2/$d.f. & $52.2/17$ & $19.7/16$ & $19.4/15$ & $20.4/16$ & $22.1/16$ & $19.7/15$ \\
\end{tabular}
\end{ruledtabular}
\end{table*}

\begin{table*}
\caption{\label{BAO_thetaMC_fit_00}
Cosmological parameters obtained from the 18 BAO measurements in Table \ref{d0}
plus $\theta_\textrm{MC}$ from Eq. (\ref{tMC}) in several scenarios.
No corrections for peculiar motions are applied.
Scenario 1 has $\Omega_\textrm{DE}$ constant.
Scenario 2 has $w(a) = w_0 + w_a (1 - a)$.
Scenario 3 has $w = w_0$.
Scenario 4 has $\Omega_\textrm{DE}(a) = \Omega_\textrm{DE} \left[1 + w_1 (1 - a)\right]$.
}
\begin{ruledtabular}
\begin{tabular}{c|cccccc} 
   & Scenario 1 & Scenario 1 & Scenario 2 & Scenario 3 & Scenario 4 & Scenario 4 \\
$\Delta d_z$ & $0.0000$ & $0.0000$ & $0.0000$ & $0.0000$ & $0.0000$ & $0.0000$ \\
\hline
$\Omega_k$ & $0$ fixed  & $0.037 \pm 0.011$ & $0$ fixed  & 
  $0$ fixed & $0$ fixed & $0.043 \pm 0.041$ \\
$\Omega_\textrm{DE} + 2 \Omega_k$ & $0.733 \pm 0.004$ & $0.716 \pm 0.006$ & $0.777 \pm 0.087$ 
  & $0.724 \pm 0.005$ & $0.718 \pm 0.006$ & $0.716 \pm 0.006$ \\
$w_0$ & n.a. & n.a. & $-0.906 \pm 0.110$ & $-0.817 \pm 0.051$ & n.a. & n.a. \\
$w_a$ or $w_1$ & n.a. & n.a. & $0.825 \pm 0.279$ & n.a. & $0.729 \pm 0.237$ & $-0.161 \pm 0.941$ \\
$d_\textrm{BAO} \times 100$ & $3.46 \pm 0.02$ & $3.38 \pm 0.03$ & $3.39 \pm 0.05$ 
  & $3.35 \pm 0.04$ & $3.34 \pm 0.04$ & $3.39 \pm 0.06$ \\
$\chi^2/$d.f. & $36.5/17$ & $23.9/16$ & $23.2/15$ & $24.3/16$ & $25.0/16$ & $23.9/15$ \\
\end{tabular}
\end{ruledtabular}
\end{table*}

\section{Cosmological parameters from BAO}

Let us try to understand qualitatively how 
the BAO distance measurements presented in Table \ref{d0}
constrain the cosmological parameters.
In the limit $z \rightarrow 0$ we obtain
$d_\textrm{BAO} = \hat{d}_\alpha(0, z_c) = \hat{d}_z(0, z_c) = \hat{d}_/(0, z_c)$,
so the row with $z = 0.1$ in Table \ref{d0}
approximately determines $d_\textrm{BAO}$.
This $d_\textrm{BAO}$ and
the measurement of, for example, $\hat{d}_z(0.3, z_c)$
then constrains the derivative of 
$\Omega_\textrm{m}/a^3 + \Omega_\textrm{DE} + \Omega_\textrm{k}/a^2$
with respect to $a$ at $z \approx 0.3$, i.e.
constrains approximately 
$\Omega_\textrm{DE} + 0.5 \Omega_k$ or equivalently 
$\Omega_\textrm{DE} - \Omega_\textrm{m}$.
The fit to the BAO data of Table \ref{d0} for Scenario 1 
(with the correction (\ref{Dd}) with $\Delta d_z = 0.0012$) obtains
$\Omega_\textrm{DE} + 0.5 \Omega_k = 0.646 \pm 0.022$, while the
constraint on the orthogonal relation is quite weak:
$\Omega_\textrm{DE} - 2.0 \Omega_k = 1.184 \pm 0.682$.
We need an additional constraint for Scenario 1.
At small $a$, $E(a)$ is dominated by $\Omega_\textrm{m}$,
so $\theta_\textrm{MC}$ plus $d_\textrm{BAO}$ approximately constrain
$\Omega_\textrm{m}$, or equivalently $\Omega_\textrm{DE} + \Omega_k$,
see Eq. (\ref{rS2}).
The additional BAO distance measurements in Table \ref{d0}
then also constrain $w_0$ and $w_a$ or $w_1$.

In Table \ref{BAO_fit} we present the
cosmological parameters	obtained by minimizing the
$\chi^2$ with 18 terms corresponding to the 
18 BAO distance measurements in Table \ref{d0} for
several scenarios. We find that the data 
(with the correction (\ref{Dd}) with $\Delta d_z = 0.0012$) is in
agreement with the simplest cosmology with
$\Omega_k = 0$ and $\Omega_\textrm{DE}(a)$ constant
with $\chi^2$ per degree of freedom (d.f.) $19.3/16$, so no additional
parameter is needed to obtain a good fit to this data.
Note that the BAO data alone place a tight constraint on 
$\Omega_\textrm{DE} + 0.5 \Omega_k = 0.646 \pm 0.022$ for
constant $\Omega_\textrm{DE}(a)$, or
$0.656 \pm 0.065$ when $\Omega_\textrm{DE}(a)$ is allowed to
depend on $a$ as in Scenario 4.
The constraint on $\Omega_k$ is weak.

In Table \ref{BAO_thetaMC_fit_12} we present the
cosmological parameters obtained by minimizing the
$\chi^2$ with 19 terms corresponding to the 
18 BAO distance measurements listed
in Table \ref{d0} 
(with the correction (\ref{Dd}) with $\Delta d_z = 0.0012$)
plus the measurement of $\theta_\textrm{MC}$
from the CMB given in Eq. (\ref{tMC}).
We present the variable $\Omega_\textrm{DE} + 2 \Omega_\textrm{k}$
instead of $\Omega_\textrm{DE}$ because it has a smaller uncertainty.
The simplest cosmology with
$\Omega_k = 0$ and $\Omega_\textrm{DE}(a)$ constant has
$\chi^2$ per d.f. $52.2/17$ and is therefore disfavored
with a significance corresponding to $4.3 \sigma$ for
one degree of freedom. Releasing one of the parameters $\Omega_k$
or $w_0$ or $w_1$ obtains a good fit to the data.
Releasing only $\Omega_k$ obtains $\Omega_k = 0.061 \pm 0.012$
with $\chi^2$ per d.f. $19.7/16$.
Releasing only $w_0$ obtains $w_0 = -0.730 \pm 0.046$
with $20.4/16$.
Releasing only $w_1$ obtains $w_1 = 1.157 \pm 0.245$
with $22.1/16$. 
Releasing both $\Omega_k$ and $w_1$ obtains $\chi^2$ per d.f. $19.7/15$,
$\Omega_k = 0.065 \pm 0.043$ and
$w_1 = -0.113 \pm 0.985$. 
Details of these
and other fits are presented in Table \ref{BAO_thetaMC_fit_12}.
Table \ref{BAO_thetaMC_fit_00} presents the corresponding
fits with no correction for peculiar motions.

In summary, we find that the BAO$+\theta_\textrm{MC}$ data are in
agreement with a family of universes with 
different $\Omega_k$. 
Two examples are
presented in Table \ref{BAO_thetaMC_fit_12}:
see the fits for Scenario 4.
Fixing $\Omega_k = 0.065$ we obtain $\Omega_\textrm{DE}$ 
independent of $a$ within uncertainties: $w_1 = -0.105 \pm 0.279$. 
Fixing $\Omega_k = 0$ we obtain
$\Omega_\textrm{DE}(a)$ that does depend significantly
on $a$: $w_1 = 1.157 \pm 0.245$.
However, if we require both $\Omega_k = 0$
and $\Omega_\textrm{DE}(a)$ constant we obtain
disagreement with the data with $\chi^2$ per d.f.
$52.2/17$.

\section{Cross-checks}

The reference fit in this Section is Scenario 1 with $\Omega_k = 0$
fixed, $\chi^2$ with 19 terms: 18 BAO terms from
Table \ref{d0} (corrected with $\Delta d_z = 0.0012$)
plus $\theta_\textrm{MC}$ from Eq. (\ref{tMC}).
The $\chi^2$ per d.f. is $52.2/17$, see Table \ref{BAO_thetaMC_fit_12}.
This tension is equivalent to $4.3 \sigma$ for one degree of freedom.
Releasing a single parameter reduces the tension
to less than $1.5 \sigma$.

The main contributions to the $\chi^2$ of the reference fit are
$\hat{d}_/(0.54, z_c)$ contributing 8.5,
$\hat{d}_z(0.46, z_c)$,
$\hat{d}_z(0.67, z_c)$,
$\hat{d}_\alpha(0.10, z_c)$,
$\hat{d}_z(0.54, z_c)$, and
$\hat{d}_\alpha(0.35, z_c)$ contributing 4.7.
So we see no obvious pattern or mistake in the
identification of the BAO signals.

Removing the term corresponding to $\theta_\textrm{MC}$
from the $\chi^2$ obtains $19.3/16$, so the tension is
between the $\theta_\textrm{MC}$ measurement and the
BAO measurements. Does $\theta_\textrm{MC}$ need a correction?

Here are fits with less BAO terms in the $\chi^2$ than the reference fit.
Fitting the 6 $\hat{d}_z(z, z_c)$ measurements plus $\theta_\textrm{MC}$ obtains $23.7/5$.
Fitting the 6 $\hat{d}_\alpha(z, z_c)$ measurements plus $\theta_\textrm{MC}$ obtains $13.7/5$.
Fitting the 6 $\hat{d}_/(z, z_c)$ measurements plus $\theta_\textrm{MC}$ obtains $11.0/5$.
Removing $\hat{d}_\alpha(0.10, z_c)$ obtains $44.7/16$.
Removing the 3 BAO entries with $z = 0.1$ obtains $39.0/14$.
Removing the 3 BAO entries with $z = 0.67$ obtains $42.4/14$.
Removing the 6 BAO entries with $z = 0.1$ and $z = 0.25$ obtains $35.3/11$.
Removing the 6 BAO entries with $z = 0.54$ and $z = 0.67$ obtains $26.0/11$.
In conclusion, none of these removals of BAO measurements from the 
$\chi^2$ of the fit removes the tension.

We now perform a fit with different BAO measurements listed in
Table \ref{d1}:
$0.2 < z < 0.4$ G-C, $0.4 < z < 0.5$ G-C, $0.5 < z < 0.6$ G-LC,
and $0.6 < z < 0.9$ LG-LG. These measurements have different
selections of centers and/or galaxies, or
different binnings of $z$, and different fits from the measurements in
Table \ref{d0}. The fit to these 12 BAO measurements
plus $\theta_\textrm{MC}$ for Scenario 1 with $\Omega_k = 0$ fixed
obtains $32.0/11$, equivalent to a tension of $3.4 \sigma$ for 1 degree of freedom.
Releasing only $\Omega_k$ obtains $4.6/10$ with $\Omega_k = 0.072 \pm 0.017$.
Releasing only $w_0$ obtains $5.9/10$ with $w_0 = -0.658 \pm 0.065$.
Releasing only $w_1$ obtains $7.2/10$ with $w_1 = 1.745 \pm 0.458$.
Note that releasing a single parameter removes the tension.
These last three fits are consistent with the ones presented in 
Table \ref{BAO_thetaMC_fit_12}, and have $\chi^2$ per d.f. less
than 1 confirming that we are not 
underestimating the systematic uncertainty of the BAO 
distances.

To remove the tension between the Scenario with $\Omega_k = 0$ fixed
and $\Omega_\textrm{DE}(a)$ constant it is necessary to shift $\theta_\textrm{MC}$
by $200 \sigma$ or increase the systematic uncertainty of the
BAO distances $\pm 0.00060$ by a factor 1.7. It is difficult to imagine
that the systematic uncertainty is wrong by a factor 1.7 when it was obtained
directly from the data, and in addition we obtain $\chi^2$ per d.f.
close to 1 by releasing a single parameter.

The conclusions are: (i) Scenario 1 with $\Omega_k = 0$
and constant $\Omega_\textrm{DE}(a)$ has tension with the BAO
plus $\theta_\textrm{MC}$ data, and (ii)
dropping the constraint $\Omega_k = 0$ and/or allowing
$\Omega_\textrm{DE}(a)$ to be variable leaves no significant tension.
These conclusions are robust with respect to the selection 
of the BAO data.

\begin{figure}
\begin{center}
\scalebox{0.45}
{\includegraphics{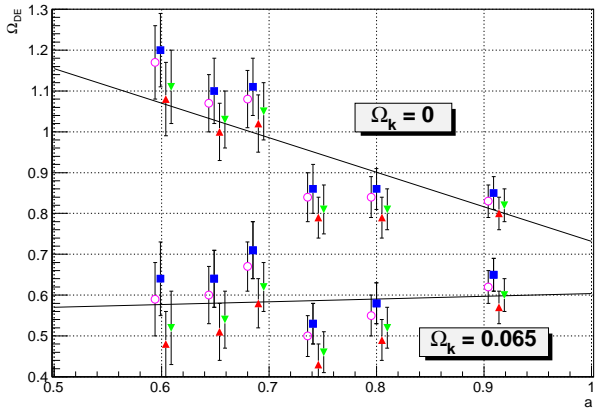}}
\caption{Measurements of
$\Omega_\textrm{DE}(a)$ obtained
from the 6 $\hat{d}_z(z, z_c)$ in Table \ref{d0} 
for $\Omega_k = 0$ and $\Omega_k = 0.065$, and
the corresponding $d_\textrm{BAO}$ and $\Omega_\textrm{DE}$
from the fits for Scenario 4 in Table \ref{BAO_thetaMC_fit_12}.
The straight lines are $\Omega_\textrm{DE} = 0.732 \left[ 1 + 1.157 (1 - a) \right]$
and $\Omega_\textrm{DE} = 0.604 \left[ 1 - 0.113 (1 - a) \right]$ from these fits.
The uncertainties correspond only to the total uncertainties
of $\hat{d}_z(z, z_c)$.
For clarity some offsets in $a$ have been applied.
For $\Omega_k =	0$ we have
$(d_\textrm{BAO}, \Omega_\textrm{DE}) = (0.0337 - 0.0004, 0.732)$ (squares),
$(0.0337 + 0.0004, 0.732)$ (triangles),
$(0.0337, 0.732 - 0.006)$ (inverted triangles),
$(0.0337, 0.732 + 0.006)$ (circles).
For $\Omega_k =	0.065$ we have 
$(0.0345 - 0.0006, 0.604)$ (squares),
$(0.0345 + 0.0006, 0.604)$ (triangles),
$(0.0345, 0.604 - 0.008)$ (inverted triangles),
$(0.0345, 0.604 + 0.008)$ (circles).
}
\label{O_DE_z}
\end{center}
\end{figure}

\begin{figure}
\begin{center}
\scalebox{0.45}
{\includegraphics{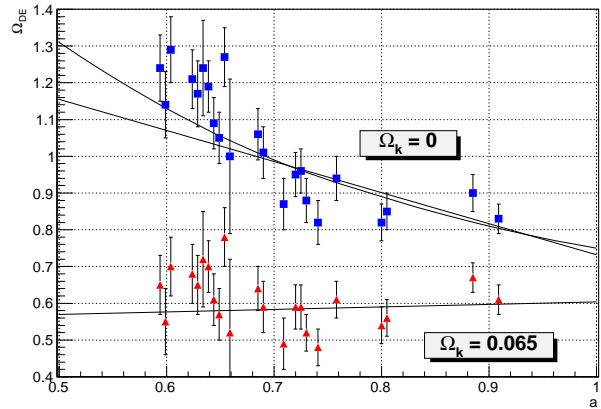}}
\caption{Same as Figure \ref{O_DE_z} with the addition of
the 17 measurements of $\hat{d}_z(z, z_c)$ in Table \ref{d1}.
These measurements are partially correlated.
For $\Omega_k = 0$ we have
$(d_\textrm{BAO}, \Omega_\textrm{DE}) = (0.0337, 0.732)$ (squares).
For $\Omega_k = 0.065$ we have
$(0.0345, 0.604)$ (triangles).
The curve corresponding to the fit for Scenario 3 in Table 
\ref{BAO_thetaMC_fit_12} has been added. 
}
\label{O_DE_z_all}
\end{center}
\end{figure}

\section{Measurement of $\Omega_\textrm{DE}(a)$}

We obtain $\Omega_\textrm{DE}(a)$ from the 6 independent measurements of
$\hat{d}_z(z, z_c)$ in Table \ref{d0} with the correction $\Delta d_z = 0.0012$
and Eqs. (\ref{dz_rs}) and (\ref{E}),
for the cases $\Omega_k = 0.065$ and $\Omega_k = 0$.
The values of $d_\textrm{BAO}$ and $\Omega_\textrm{DE}$
are obtained from the fits for Scenario 4 in Table \ref{BAO_thetaMC_fit_12}.
The results are presented in Fig. \ref{O_DE_z}. To guide the eye, we
also show the straight line corresponding to Scenario 4.

To cross-check the robustness of $\Omega_\textrm{DE}(a)$ in Fig. \ref{O_DE_z}
we add the 17 measurements of $\hat{d}_z(z, z_c)$ in Table \ref{d1} and
obtain Fig. \ref{O_DE_z_all}. Note that these measurements of
$\hat{d}_z(z, z_c)$ are partially correlated.

Note that $\Omega_\textrm{DE}(a)$ is consistent with a
constant for $\Omega_k = 0.065$, but not for $\Omega_k = 0$.

\section{Conclusions}

The main results of these studies are the 18 independent
BAO distance measurements presented in Table \ref{d0}
which do not depend on any cosmological parameter.
These BAO distance measurements alone place a strong
constraint on $\Omega_\textrm{DE} + 0.5 \Omega_k = 0.646 \pm 0.022$
(for constant $\Omega_\textrm{DE}$)
or $0.656 \pm 0.065$ (when $\Omega_\textrm{DE}(a)$ is allowed to depend on $a$),
while the constraint on $\Omega_k$ is weak.

Constraints on the cosmological parameters
in several scenarios from the BAO measurements alone
and from BAO plus $\theta_\textrm{MC}$ measurements are presented in
Tables \ref{BAO_fit}, \ref{BAO_thetaMC_fit_12} and \ref{BAO_thetaMC_fit_00}.
We find that the 18 BAO distance measurements
plus $\theta_\textrm{DE}$ are in agreement with a family of universes
with different $\Omega_k$. Two examples are
$\Omega_k = 0.065$ with $\Omega_\textrm{DE}(a)$ constant,
and $\Omega_k = 0$ with $\Omega_\textrm{DE}(a)$ 
varying significantly with $a$.
For these two examples we present measurements of
$\Omega_\textrm{DE}(a)$ in 
Figs. \ref{O_DE_z} and \ref{O_DE_z_all}.
The cosmology with both $\Omega_k = 0$ and
$\Omega_\textrm{DE}(a)$ constant has a tension of
$4.3 \sigma$ with the BAO plus $\theta_\textrm{MC}$ data. 

The BAO plus $\theta_\textrm{MC}$ data for constant
$\Omega_\textrm{DE}(a)$ 
obtains $\Omega_k = 0.061 \pm 0.012$ for $\Delta d_z = 0.0012$
and $\Omega_k = 0.037 \pm 0.011$ for $\Delta d_z = 0$.
These results have some tension with
independent observations:
$\Omega_k = -0.042^{+0.024}_{-0.022}$ from
CMB data alone with the assumption of constant
$\Omega_\textrm{DE}(a)$ \cite{PDG}, and
$\Omega_k \approx 0.000 \pm 0.013$ from
CMB plus supernova (SN) data with the assumption of constant
$\Omega_\textrm{DE}(a)$ \cite{PDG}.
If $\Omega_\textrm{DE}(a)$ is
allowed to be variable
there are only loose constraints on $\Omega_k$
from independent measurements, so
the case with $\Omega_k = 0$ with non-constant
$\Omega_\textrm{DE}(a)$ is viable.

\section{Acknowledgment}
Funding for SDSS-III has been provided by the 
Alfred P. Sloan Foundation, the Participating Institutions, 
the National Science Foundation, and the 
U.S. Department of Energy Office of Science. 
The SDSS-III web site is http://www.sdss3.org/.

SDSS-III is managed by the Astrophysical Research Consortium 
for the Participating Institutions of the SDSS-III Collaboration 
including the University of Arizona, the Brazilian Participation Group, 
Brookhaven National Laboratory, Carnegie Mellon University, 
University of Florida, the French Participation Group, 
the German Participation Group, Harvard University, 
the Instituto de Astrofisica de Canarias, 
the Michigan State/Notre Dame/JINA Participation Group, 
Johns Hopkins University, Lawrence Berkeley National Laboratory, 
Max Planck Institute for Astrophysics, Max Planck Institute for Extraterrestrial Physics, 
New Mexico State University, New York University, Ohio State University, 
Pennsylvania State University, University of Portsmouth, Princeton University, 
the Spanish Participation Group, University of Tokyo, 
University of Utah, Vanderbilt University, University of Virginia, 
University of Washington, and Yale University. 

The author acknowledges the use of computing resources of
Universidad de los Andes, Bogot\'{a}, Colombia.

\end{document}